%% file: manuscript.tex
\documentclass[twocolumn,aps,prl,preprintnumbers,superscriptaddress]{revtex4-2}
\usepackage[utf8]{inputenc}
\usepackage{amsmath}
\usepackage{amssymb}
\usepackage{graphicx}
\usepackage{textcomp}% Include figure files
\usepackage{bm}% bold math
\usepackage[right]{eurosym}
\usepackage{float}
\usepackage[english]{babel}
\usepackage{blindtext}
\usepackage{booktabs}
\usepackage{longtable}
\usepackage{tikz}
\usepackage{siunitx}
\usepackage{placeins}
\newcommand{\ICD}{{\small ICD}}
\newcommand{\AI}{{\small AI}}
\newcommand{\PI}{{\small PI}}
\newcommand{\PL}{{\small PL}}
\newcommand{\EEL}{{\small EEL}}
\newcommand{\VMI}{{\small VMI}}
\newcommand{\HEA}{{\small HEA}}
\newcommand{\SM}{{\small SM}}
\newcommand{\ASTRID}{{\small ASTRID2}}

\begin{document}

\author{B. Bastian}
\affiliation{Wilhelm Ostwald Institute for Physical and Theoretical Chemistry, University of Leipzig, Linn\'{e}stra\ss{}e 2, 04103 Leipzig, Germany}
\author{J. D. Asmussen}
\author{L. Ben Ltaief}
\author{H. B. Pedersen}
\affiliation{Department of Physics and Astronomy, Aarhus University, 8000 Aarhus C, Denmark}
\author{K. Sishodia}
\author{S. De}
\affiliation{Indian Institute of Technology Madras, Chennai 600036, India}
\author{S. R. Krishnan}
\affiliation{Department of Physics and QuCenDiEM-group, Indian Institute of Technology Madras, Chennai 600036, India}
\author{C. Medina}
\affiliation{Institute of Physics, University of Freiburg, 79104 Freiburg, Germany}
\author{N. Pal}
\author{R. Richter}
\affiliation{Elettra-Sincrotrone Trieste, 34149 Basovizza, Trieste, Italy}
\author{N. Sisourat}
\affiliation{Sorbonne Universit\'e, CNRS, Laboratoire de Chimie Physique Mati\`ere et Rayonnement, UMR 7614, F-75005 Paris, France}
\author{M. Mudrich}\email{mudrich@phys.au.dk}
\affiliation{Department of Physics and Astronomy, Aarhus University, 8000 Aarhus C, Denmark}

\title{Observation of interatomic Coulombic decay induced by double excitation of helium in nanodroplets}
\begin{abstract}
Interatomic Coulombic decay (\ICD{}) plays a crucial role in weakly bound
complexes exposed to intense or high-energy radiation. So far, neutral or ionic
atoms or molecules have been prepared in singly excited electron or hole states which can
transfer energy to neighboring centers and cause ionization and
radiation damage. Here we demonstrate that a doubly excited atom,
despite its extremely short lifetime, can decay by \ICD{}; evidenced
by high-resolution photoelectron spectra of He nanodroplets excited
to the 2s2p+ state. We find that
\ICD{} proceeds by relaxation into excited He$^*$He$^+$ atom-pair states, in
agreement with calculations. The ability of inducing \ICD{} by
resonant excitation far above the single-ionization threshold opens
opportunities for controlling radiation damage to a high degree of element
specificity and spectral selectivity.
\end{abstract}
\date{\today}
\maketitle

Highly excited atoms and molecules embedded in an environment can efficiently decay by exchange of energy or charge with an atom or molecule in their environment through interatomic (intermolecular) electronic correlation. These processes, termed interatomic (intermolecular) Coulombic decay (\ICD{}), cause ionization of the neighboring center, thereby quenching concurrent decay processes~\cite{jahnke_ultrafast_2010,FORSTEL201316,jabbari2020competition,hergenhahn_interatomic_2011,Jahnke15:082001,Jahnke20}. Since the prediction of \ICD{} as an efficient decay process of clusters with an excited intermediate-shell electron~\cite{Cederbaum1997}, a variety of related processes have been discovered involving both neutral and ionic species. The general relevance of \ICD{} has been established, with implications for radiation damage of biological matter ~\cite{Stoychev11:6817,Gokhberg2014,stumpf2016role,Ren2018,Gopakumar:2023}.

\begin{figure}
  \begin{center}
    \includegraphics[width=1\columnwidth]{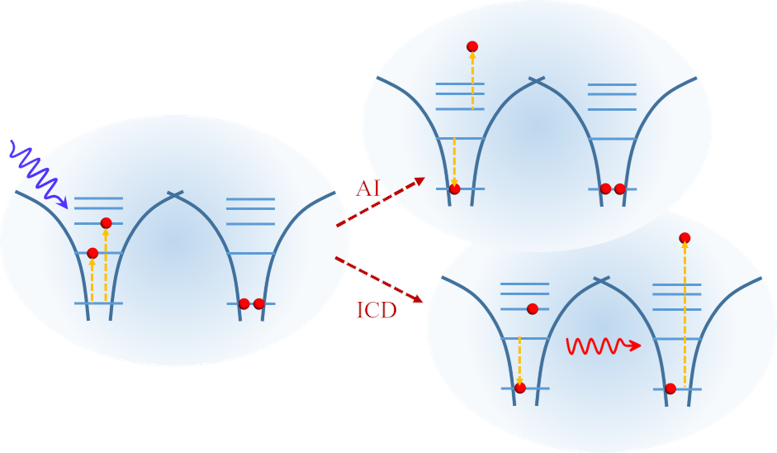}
    \caption{Schematics of a pair of He atoms where one is in a doubly excited state that can decay by \textit{intra}atomic autoionization (\AI{}) or by \textit{inter}atomic Coulombic decay (\ICD{}).
    \label{fig:scheme}}
  \end{center}
\end{figure}

A special case of electronic excitation is doubly excited states of atoms which manifest themselves as resonances in photoionization (\PI{}) spectra. A paradigm system is the helium atom (He$^{**}$) where the two electrons occupy discrete excited states within the photoionization continuum. Excitation into these states leads to the same final state via two pathways; direct \PI{} and autoionization (\AI{}) of the He$^{**}$ both form a He$^+$ and an electron. Quantum interference of the two indistinguishable paths causes the well-known asymmetric Fano line shape~\cite{fano_effects_1961}. The first series of these Fano resonances appears around 60~eV photon energy~\cite{Cooper1963518,Domke:1996,rost1997resonance}. 
When a second groundstate He atom is placed next to the He$^{**}$, an additional non-local autoionization process may open up\,---\,\ICD{}; see Fig.~\ref{fig:scheme} for a schematic representation.
The decay of isolated He$^{**}$ atoms (upper pathway) has been studied in detail both theoretically and experimentally; the lowest optically accessible doubly excited state, 2s2p+, decays by \AI{} in only 17.5~fs~\cite{Cooper1963518,Domke:1996}. Decay of He$^{**}$ by \ICD{} (lower path) has only been considered theoretically for the diatomic system He$^{**}$Ne~\cite{jabbari2020competition}. In this process, one of the two electrons decays to the ground state and the resulting energy is transferred to a neighboring atom which in turn is ionized. The decay by fluorescence emission only becomes a relevant concurrent decay path for higher states and is therefore disregarded here.

\ICD{} of doubly excited atoms is expected to be particularly efficient because these states are more sensitive to perturbations by neighboring atoms than those where only a single electron is excited~\cite{jabbari2020competition}. In a more complex environment, which has not been considered until now, this type of \ICD{} may be even faster; generally the \ICD{} rate steeply rises for decreasing interatomic distance between the excited and the neighboring atom and it scales approximately proportionally to the number of neighbors~\cite{Cederbaum1997,kuleff_ultrafast_2010}.

Here we present a combined experimental and theoretical study of \ICD{} of doubly excited He atoms in He nanodroplets. He nanodroplets have proven to be particularly well suited test beds for exploring \ICD{} processes due to the simple electronic structure of the He atom which makes electron spectra easy to interpret and theoretical calculations tractable. The homogeneous, superfluid density distribution of He nanodroplets and their property of picking up virtually any foreign atom or molecule have enabled us to study various types of \ICD{} in homogeneous and heterogeneous systems such as metal-doped He nanodroplets~\cite{LaForge2016,Shcherbinin2017,LaForge2019,BenLtaief2019,Ltaief2020,Ovcharenko2020,LaForge2021,asmussen2022time,ltaief2023efficient}. Double-excitation spectra of He nanodroplets feature Fano profiles with similarly asymmetric shape as the atomic lines. However, the droplet profiles are blue shifted and significantly broadened~\cite{laforge_fano_2016}.

In this work we use two different techniques to measure photoelectron spectra: (i) A velocity-map imaging (\VMI{}) spectrometer located at the synchrotron radiation source \ASTRID{} at Aarhus University allows us to detect electrons and ions in coincidence
~\cite{bastian2022new}. (ii) A hemispherical electron analyzer (\HEA{}) used at the Gasphase beamline of Elettra synchrotron in Trieste (model VG 220i mounted at the magic angle) allows us to measure high-resolution electron spectra.
With the \VMI{} spectrometer we detect the full distribution of electron energies $E_e$ at photon energies in the range $h\nu=58$\,--\,63~eV.
\VMI{} data were recorded at different helium expansion conditions with a stagnation pressure of 30~bar and nozzle temperatures from 10~K to 16~K corresponding to average droplet sizes from $1.4\times 10^6$ to $7.5\times 10^3$ atoms per droplet as determined by titration measurements~\cite{Gomez2011}.
In the \VMI{} measurements at 12~K ($4.3\times 10^4$ atoms), the droplets have comparable
sizes as in the \HEA{} measurements at 14~K and 50~bar~\cite{Toennies2004}.
In both setups, background and foreground data were measured intermittently using a chopper wheel that periodically blocks and unblocks the droplet beam.

To compute the total and partial decay widths of the He(2s2p)--He(1s$^2$)
$^1\Sigma_u$ and $^1\Pi_u$ states, we used the R-Matrix method~\cite{Tennyson2010} as implemented in
the UKRmol+ package~\cite{ukrmol+}.
We used Complete Active Space Self Consistent Field (CASSCF) molecular
orbitals, computed with the MOLPRO package~\cite{molpro1,molpro2}. In our
CASSCF, 18 orbitals were used and all electrons were active. The orbitals were
optimized for the ground electronic state of He$_2^+$. We performed the
calculations with the aug-cc-pVDZ basis set. In the scattering calculations, we
employed a Close-Coupling (CC) model in which the lowest 6 electronic states of
the target He$^*$He$^+$ are included. The continuum-like orbitals are
described with 25 B-splines at each order, up to the order 6. The R-matrix
radius was fixed at 6.88~\AA{}. For the outer region calculations, the R-matrix
is propagated from 6.88~\AA{} to 42~\AA{}. The maximum multipole to be
retained in the expansion of the long range potential was set to 2. The total
resonance widths and partial widths have been obtained with the programs
RESON~\cite{reson} and TIMEDEL~\cite{timedel}.
Convergence of the decay widths with respect to the R-matrix parameters and the
CC model was checked. The UKRmol-in-2.0.2 and UKRmol-out-2.0.1 release versions
of the UKRmol+ package were used.

\begin{figure}
  \begin{center}
    \includegraphics[width=0.85\columnwidth]{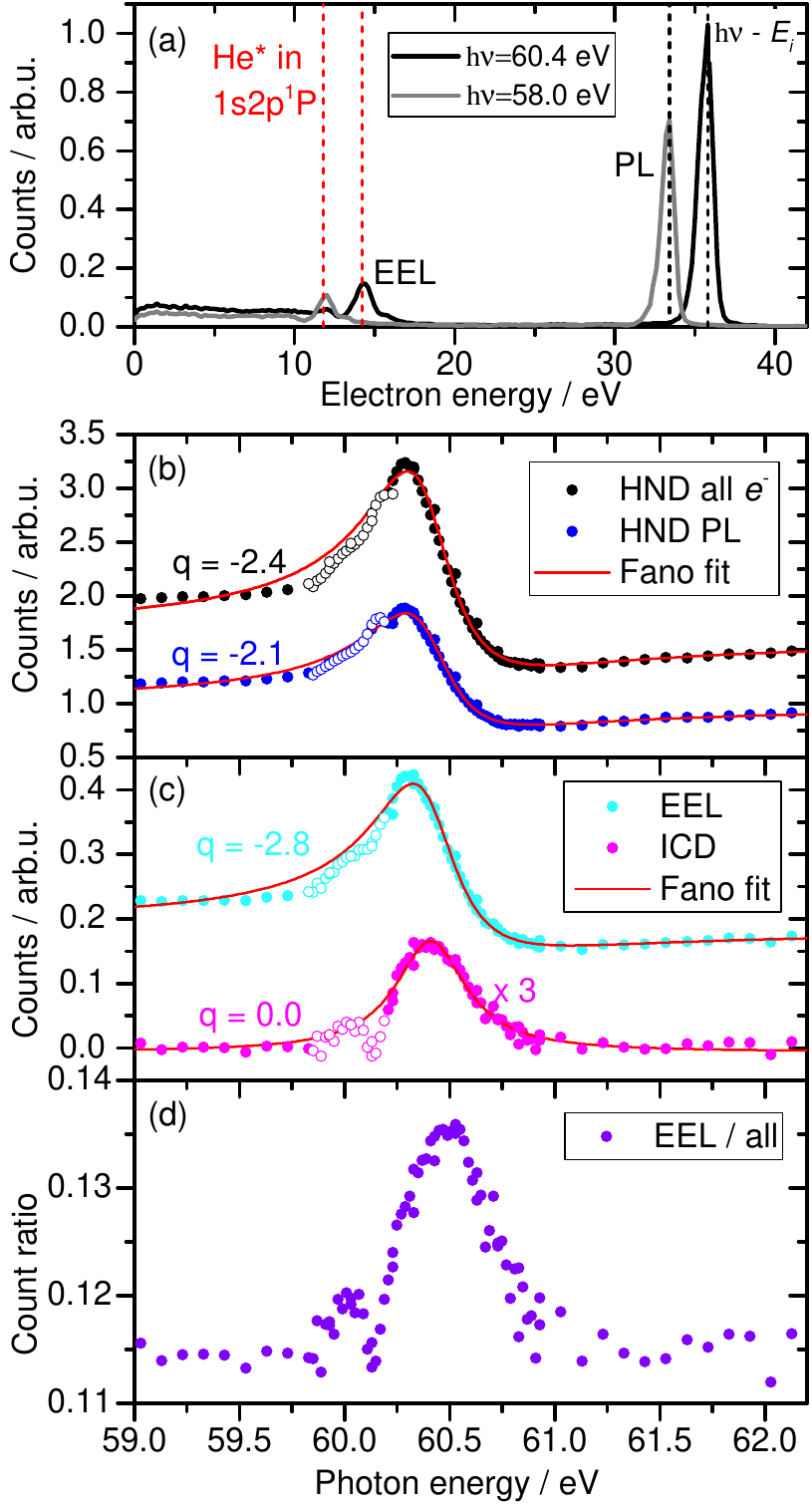}
    \caption{Photoionization spectra of He nanodroplets ($\langle N\rangle\approx 4.3 \times 10^4$ He atoms per droplet) around the 2s2p+ Fano resonance.
      \textbf{(a)} Photoelectron spectra recorded at $h\nu=58$~eV and 60.4~eV using the \VMI{} spectrometer.
      \textbf{(b)} Yield of direct photoelectrons and total electrons as a function of photon energy.
      \textbf{(c)} Yield of electrons subjected to energy loss due to inelastic scattering or \ICD{}.
      \textbf{(d)} The ratio of energy loss to total electrons features a pronounced maximum indicating \ICD{} at the resonance.
      For fits of the Fano profile in (b) and (c), only full circles were considered to exclude the sharp helium atomic line due to the atomic component in the droplet beam.
    \label{fig:Ratio}}
  \end{center}
 \end{figure}

The electron spectra inferred from the \VMI{} measurements contain two prominent features, see Fig.~\ref{fig:Ratio}~(a). A sharp peak at $E_e=h\nu-E_i$ electron kinetic energy, denoted as photoline (\PL{}), is due to direct \PI{} and \AI{} which are indistinguishable processes. Here, $E_i=24.59~$eV is the ionization energy of He. The small peak around $E_e=15$~eV results mainly from electron energy loss (\EEL{}) by inelastic scattering of the outgoing photoelectron at He atoms in the nanodroplet surrounding the photoion, according to the reaction~\cite{shcherbinin2019inelastic}
\[
\mathrm{He}_N+h\nu\rightarrow \mathrm{He}^+\mathrm{He}_{N-1} + e^-_\mathrm{\PL{}}\rightarrow \mathrm{He}^+\mathrm{He}^*\mathrm{He}_{N-2} + e^-_\mathrm{\EEL{}}.
\]
In addition, \ICD{} may contribute to this feature,
\[
\mathrm{He}_N+h\nu\rightarrow \mathrm{He}^{**}\mathrm{He}_{N-1}\rightarrow \mathrm{He}^*\mathrm{He}^+\mathrm{He}_{N-2} + e^-_\mathrm{\ICD{}}.
\]
As the two reactions yield the same final state up to different 1s$n\ell$ levels of He$^*$, they are hard to distinguish in the electron spectra;
see also Supplemental Material (\SM{}) Fig.~1~\cite{sm}.
\nocite{hertel2008atoms,Mezei2015:epjw}%
However, \ICD{} can only occur when the photon energy is tuned to a He$^{**}$ resonance. Thus, we identify the \ICD{} contribution by analyzing the yields of \EEL{} electrons and total electrons as a function of photon energy $h\nu$.
For each value of $h\nu$, the \EEL{} spectra given by $E_l=h\nu-E_i-E_e$ are integrated from $-$2 to 2~eV for the photoelectrons, from 18 to 23~eV for the \EEL{}, and over the full range to obtain the total electron yield.

Fig.~\ref{fig:Ratio}~(b) shows the yield of electrons emitted from He nanodroplets by \PI{} and \AI{} and (c) shows the yield of \EEL{} electrons. The ratio of the two signals, shown in (d), reflects the probability of forming He$^*$; it features a maximum at $h\nu = 60.41\pm{}0.01~$eV, coinciding with the resonance energy that lies in the falling edge of the Fano profile.
The relative contribution to this feature from electron-He inelastic scattering is not expected to significantly vary in the narrow tuning range $h\nu=60$\,--\,61~eV around the He$^{**}$ double excitation~\cite{ralchenko2008electron}.
The resonant increase of the ratio compared to its average for off-resonant photon energies $h\nu<$~59.9~eV is therefore attributed to \ICD{}.
From the difference to the off-resonant ratio, we estimate the \ICD{} electron yield shown in Fig.~\ref{fig:Ratio}~(c), see \SM{} Eq.~(2)~\cite{sm}.
Interestingly, the spectral-line shapes in Fig.~\ref{fig:Ratio}~(b-c) clearly differ from one another. Fits of the standard Fano profile~\cite{Domke:1996,laforge_fano_2016} to the data (solid lines) result in an asymmetry parameter (or Fano parameter) $q=-2.13 \pm 0.05$ for the photoline and $q=-2.36 \pm 0.07$ for all electrons.
The latter coincides with a previous measurement of all emitted electrons
(mostly photoelectrons)~\cite{laforge_fano_2016} and comes close to the
value for the atomic Fano profile ($q=-2.77$,~\cite{Domke:1996}).
For the \EEL{} profile we find a similar value $q=-2.8 \pm 0.1$, while the \ICD{} profile yields $q=0.00 \pm 0.02$.
Note that for $q=0$ the Fano resonance profile is identical to the Breit–Wigner or Lorentzian formula for resonant excitation of a two-level system.
The \ICD{} signal is expected to follow the latter resonance curve because \ICD{} can
only occur for the one path involving He$^{**}$ excitation, whereas the path
of direct photoionization cannot lead to \ICD{}.
Symmetrically broadened \ICD{} lines were previously observed for
inner-valence-shell excited rare-gas dimers and clusters~\cite{Flesch2014:zpc}.

\begin{figure}
  \begin{center}
    \includegraphics[width=\columnwidth]{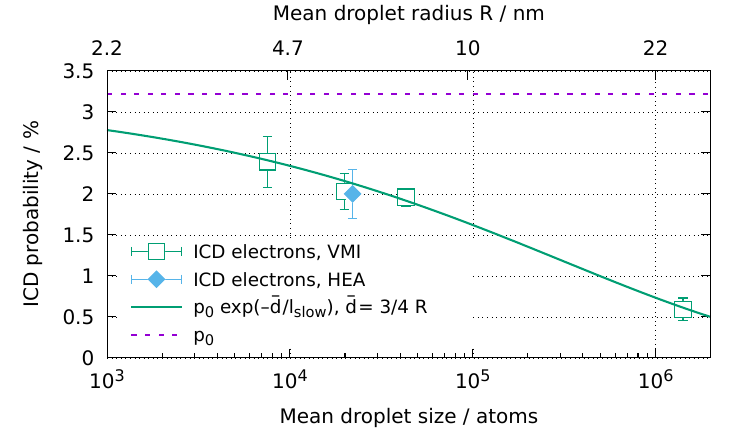}
    \caption{Probabilities for the detection of \ICD{} electrons relative to the total number of electrons as a function of average droplet size. The \HEA{} point is computed as product of the \ICD{} to \EEL{} ratio and the \EEL{} probability for \VMI{} data at a comparable average droplet size.
      The \VMI{} data is fitted with a simple model to estimate
      the nascent \ICD{} probability $p_0$.
    \label{fig:probabilities}}
  \end{center}
\end{figure}

The \VMI{} measurements at different expansion temperatures allow
us to analyze \ICD{} probabilities relative to total
electrons as a function of droplet size (see Eq.~(3), Sec.~III and Tab.~I
in the \SM{}~\cite{sm}).
As \ICD{} mostly occurs between nearest neighbors, no pronounced size dependence is
expected in the studied range of droplet sizes; the apparent drop of the \ICD{}
probability in Fig.~\ref{fig:probabilities}
for large droplets is attributed to the trapping of slow \ICD{} electrons in bubbles that remain transiently bound to large droplets~\cite{asmussen2023dopant}.
To quantify this effect, we assume a simple model, $p_{\rm \ICD{}} = p_0 \exp\left(-\bar d/l_{\rm slow}\right)$, where $\bar d$ is the mean distance of the \ICD{} electron to the droplet surface (see \SM{} Eqs.~(9--12)~\cite{sm}) and $l_{\rm slow}$ is a mean free path.
The fit to the data, shown as solid red line in Fig.~\ref{fig:probabilities}, yields $l_{\rm slow} = (11.1 \pm 1.1)$~nm and $p_0 = (3.2 \pm 0.2)$\,\%
as the nascent \ICD{} probability inside the droplet. Extrapolation to larger droplets
shows that only one per thousand \ICD{} electrons will be ejected out of a droplet
of radius 100~nm which roughly matches the photoelectron trapping range of
150~nm for 15~eV electrons obtained from simulations, see the \SM{}
of~\cite{asmussen2023electron}.
\ICD{} and electron impact excitation probabilities for total electrons and
electron--ion coincidences with He$_{n}^{+}$ ($n=1,2,3$) as a function of
droplet size are presented in \SM{} Sec.~V~\cite{sm}.

Our theoretical computations give a rough estimate of the
\ICD{} probability in competition with \AI{} in the He$^{**}$He system
for the $^1\Sigma_u^+$ and $^1\Pi_g$ doubly excited states at five different
pair distances $r$ from 2.1 to 3.2~\AA{}, respectively.
As detailed in \SM{} Sec.~VII~\cite{sm}, we fit the average probability
with an exponential function $p(r)$
and weight it with the pair-distance distribution function
$g(r)$~\cite{Chin1995:prb,peterka2007photoionization}
to obtain the overall {\small \ICD{}}
probability relative to the total ionization rate in the He droplet environment
\begin{align*}
  p_0^{\rm theory}
  = f_{\rm res} \int p(r) g(r) \,{\rm d} r
  = (1.6 \pm 1.4)\,\%
\end{align*}
where $f_{\rm res} = 0.51 \pm 0.03$ is the fraction of double excitation
relative to total ionization at the resonance energy $E_r = 60.4$~eV,
inferred from fits of the Fano profile (see \SM{} Sec.~VII~\cite{sm}).
The theoretical estimate is based on calculations for a helium dimer.
Many-body effects of the surrounding helium atoms are therefore neglected which may
explain the difference to the somewhat larger experimental value.
Increased \ICD{} to \AI{} ratios are expected for higher-lying doubly excited states
but preliminary results in \SM{} Fig.~3~(e),(f)~\cite{sm} cannot discern the
expected resonance in the \EEL{} to \PL{} ratio from the noise level.

\begin{figure}
  \includegraphics[width=.9\columnwidth]{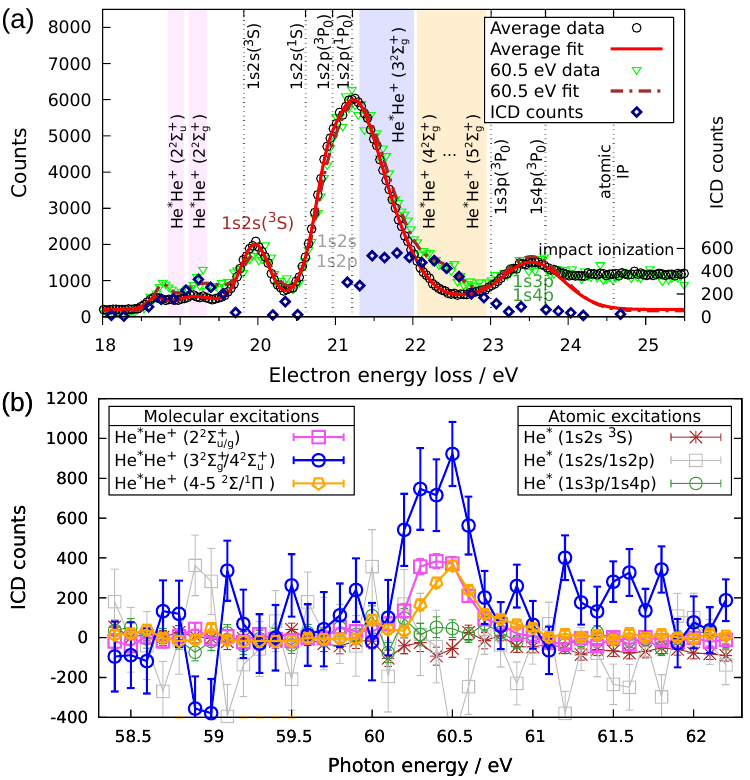}\\[-3pt]
  \includegraphics[width=.9\columnwidth]{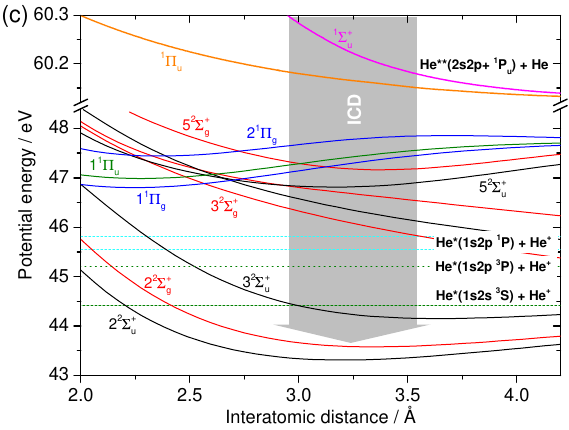}\\
  \begin{center}
    \caption{%
    \textbf{(a)} \HEA{} high-resolution \EEL{} spectrum. Black circles are
    averaged over 39 spectra recorded in the $h\nu=58.4$\,--\,62.2~eV range.
    Green triangles show a spectrum at $h\nu=60.5$~eV.
    The solid red and dash-dotted brown lines show the respective multi-peak
    fits of a sum of 8 Gaussians.
    Shaded areas indicate regions where He$^{**}$ decays by \ICD{} 
    into states shown in (c).
    Vertical dashed lines annotate atomic excitations with values
    from NIST~\cite{nist13}.
    \textbf{(b)}~Photon energy dependence of the total \ICD{} counts for excitation
    of different molecular and atomic excitations.
    \textbf{(c)}~Potential energy curves of He$_2$ after double excitation of one atom and excited states of the He$_2^+$ molecular ion.
    The color encodes different symmetries as indicated by the labels.
    Horizontal dotted lines give the asymptotes for excitation of distant helium atoms by inelastic scattering. The gray area indicates the range of distances between neighboring He atoms in He nanodroplets where \ICD{} takes place~\cite{peterka2007photoionization}.
    The potential energy curves (PECs) of He$^{**}$He states
    were taken from Ref.~\cite{laforge_fano_2016}, while the PECs of He$^*$He$^+$ states
    were computed using the full configuration interaction approach, as implemented
    in MOLPRO, with the aug-cc-pVDZ basis set.
    \label{fig:EEL}}
  \end{center}
\end{figure}

To obtain more detailed insights into the \ICD{} process we recorded
high-resolution electron spectra of the \EEL{} feature using the \HEA{}.
Fig.~\ref{fig:EEL}~(a) shows the average of 39 spectra recorded in the range
$h\nu=58.4$\,--\,62.2~eV (black circles) and a spectrum on resonance
($h\nu=60.5$~eV, green triangles). Certain regions of the spectrum around 19 and
22.5~eV are clearly enhanced on resonance. This is confirmed by multi-peak fits
consisting of the sum of 8 Gaussians to account for various final states
populated either by impact excitation by the photoelectron or by \ICD{}.
The peak widths, except that for the 1s2s(\textsuperscript{3}S) state, and the positions of the three Gaussian functions within the dominant 1s2p peak were fixed to fit the spectra for each energy, with their values determined from the
well converged fit of the average spectrum
(individual peaks and fit residuals are shown in \SM{} Fig.~3~(a)~\cite{sm}).
For each state and photon energy,
the \EEL{}-to-total-electron ratio in \SM{} Fig.~3~(b)~\cite{sm}
is computed to estimate the \ICD{} fraction
and total counts, see \SM{} Sec.~I/II~\cite{sm}.

The \ICD{} counts are presented in Fig.~\ref{fig:EEL}~(b).
The atomic He$^*$ components do not contribute to the \ICD{} counts
and are solely caused by electron impact excitation. The same is true for the electron impact ionization component, see \SM{} Fig.~3~(c)~\cite{sm}.
Impact excitation of a He$^*$ atom in a He droplet by the photoelectron mostly
occurs at some distance from the He$^+$ photoion (see
\cite{shcherbinin2019inelastic} and \SM{} Sec.~VI~\cite{sm}). In contrast, \ICD{}
produces a He$^*$He$^+$ atom-pair state at short interatomic distance given by
the spacing between neighboring He atoms in droplets.
This explains that \ICD{} counts are
exclusively observed for He$^*$He$^+$ molecular product states.

The relevant potential energy curves in this range are shown in
Fig.~\ref{fig:EEL}~(c). The gray arrow indicates the range of typical distances
from 3.0~\AA{} (most probable next-neighbor distance) to 3.6~\AA{} (average
He--He distance in He droplets)~\cite{peterka2007photoionization}. The decay
from He$^{**}$ to the lowest lying $2\,^2\Sigma_u^+$ and $2\,^2\Sigma_g^+$ as well as a
range of higher He$^*$He$^+$ states match the resonant regions in the electron
energy loss spectrum in Fig.~\ref{fig:EEL}~(a).
The large peak around 21.3~eV consists mainly of non-resonant 1s2s and 1s2p-correlated atomic excitations which are shifted and broadened in droplets~\cite{shcherbinin2019inelastic} and thus overlap with the resonant
$3\,^2\Sigma_g^+$ and $4\,^2\Sigma_u^+$ components caused by \ICD{}.
The higher lying $^2\Sigma/^1\Pi$ states can energetically mix;
we refrain from a more detailed assignment of states in this region due to
congestions of electron spectra and the limited accuracy of the  potential curves.
Increasing photon energies favor higher levels of excitation
(see \SM{} Fig.~3~(d)~\cite{sm}).

In conclusion, we presented the first experimental evidence for \ICD{} involving a doubly excited atom.
To disentangle this process from photoelectron impact excitation of an atom in the He nanodroplet,
high-resolution electron spectra were recorded at different photon energies across the
double excitation resonance of the 2s2p+ state.
\ICD{} manifests itself as enhanced yields of He$^*$He$^+$ atom pairs and
corresponding electron spectra at the Fano resonance.
From the droplet-size dependence we estimate an \ICD{} probability
relative to the total ionization rate of $(3.2 \pm 0.2)$\,\%
in decent agreement with theoretical results.
Consequently, \ICD{} competes with \AI{} despite the ultrashort decay time
(17.5~fs~\cite{Cooper1963518,Domke:1996}) of the latter.
For higher-lying doubly excited states, the \ICD{} probability is expected to
further increase relative to \AI{}~\cite{jabbari2020competition}. This type of
\ICD{} could be relevant in other systems such as water~\cite{kato2004doubly}
and solvated metal atoms~\cite{PhysRevA.31.250,baig1996inner} and thus
present a way to site-selectively deposit energy in condensed media
such as biological tissue~\cite{Stumpf2016,Gopakumar:2023}.

\section*{Acknowledgements}
We gratefully acknowledge financial support by the Danish Council for Independent Research (Grant No. 1026-00299B), the Carlsberg Foundation, the German Science Foundation (DFG) through project STI 125/19-2, and the Danish Agency for Science, Technology, and Innovation through the instrument center DanScatt. S.R.K. gratefully acknowledges the financial support from the Min.\ of Education, Govt.\ of India through the scheme for promotion of research and academic collaboration and the Institute of Eminence programmes, Govt.\ of India, and from the Max Planck Society, Germany.  S.R.K., S.D. and K.S. thank for the support by the Indo-French Center for Promotion of Academic Research (CEFIPRA), Deutsche Akademischer Austauschdienst Dienst (DAAD), Department of Science and Technology, Govt.\ of India and the Indo-Elettra scheme.  The research leading to these results has been supported by the project CALIPSOplus under grant agreement 730872 from the EU Framework Programme for Research and Innovation HORIZON 2020 and by the COST Action CA21101 ``Confined Molecular Systems: From a New Generation of Materials to the Stars (COSY)''. M.M. and J.D.A. thank Lars B. Madsen for fruitful discussions. 

\bibliography{Bib}

\end{document}

\author{B. Bastian}
\affiliation{Wilhelm Ostwald Institute for Physical and Theoretical Chemistry, University of Leipzig, Linn\'{e}stra\ss{}e 2, 04103 Leipzig, Germany}
\author{J. D. Asmussen}
\author{L. Ben Ltaief}
\author{H. B. Pedersen}
\affiliation{Department of Physics and Astronomy, Aarhus University, 8000 Aarhus C, Denmark}
\author{K. Sishodia}
\author{S. De}
\affiliation{Indian Institute of Technology Madras, Chennai 600036, India}
\author{S. R. Krishnan}
\affiliation{Department of Physics and QuCenDiEM-group, Indian Institute of Technology Madras, Chennai 600036, India}
\author{C. Medina}
\affiliation{Institute of Physics, University of Freiburg, 79104 Freiburg, Germany}
\author{N. Pal}
\author{R. Richter}
\affiliation{Elettra-Sincrotrone Trieste, 34149 Basovizza, Trieste, Italy}
\author{N. Sisourat}
\affiliation{Sorbonne Universit\'e, CNRS, Laboratoire de Chimie Physique Mati\`ere et Rayonnement, UMR 7614, F-75005 Paris, France}
\author{M. Mudrich}\email{mudrich@phys.au.dk}
\affiliation{Department of Physics and Astronomy, Aarhus University, 8000 Aarhus C, Denmark}

\title{Supplemental material\\[8pt]
  Observation of interatomic Coulombic decay induced \\
  by double excitation of helium in nanodroplets
  }
\date{\today}
\maketitle
\onecolumngrid

\begin{figure}[ht]
  \begin{center}
    \includegraphics[width=0.5\columnwidth]{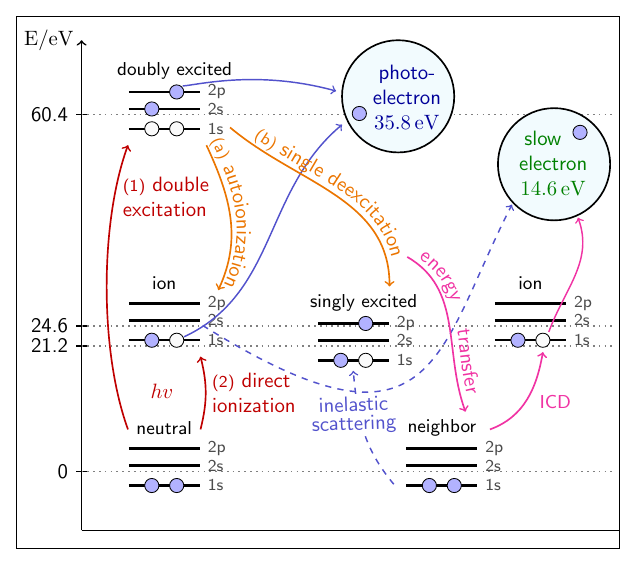}
    \newcommand{\one}[1]{{\color{red!75!black}#1}}
    \newcommand{\two}[1]{{\color{orange!92!black}#1}}
    \newcommand{\el}[1]{{\color{blue!60!white!80!black}#1}}
    \newcommand{\icd}[1]{{\color{magenta!80!white}#1}}
    \newcommand{\slow}[1]{{\color{green!50!black}#1}}
    \caption{%
      Interactions at a double excitation (Fano) resonance in a helium nanodroplet.
      An incident photon \one{$h\nu$} can \one{(1)}~doubly excite a helium atom or \one{(2)}~directly ionize. A doubly excited helium atom will quickly autoionize~\two{(1a)} or undergo single deexcitation~\two{(1b)}.
      The autoionization~\two{(1a)} and direct ionization~\one{(2)} pathways are indistinguishable and give rise to the Fano resonance profile when varying the photon energy. The resulting photoelectron may \el{escape} the droplet
      with \SI{35.8}{eV} kinetic energy or excite a neighbor atom by \el{inelastic scattering} (dashed) and escape as a slow \slow{\SI{14.6}{eV}} electron. In the single deexcitation pathway~\two{(1b)},
      instead the doubly excited atom decays to a singly excited one while a neigboring atom is ionized
      by \icd{energy transfer} and a \slow{\SI{14.6}{eV}} {\small ICD}~electron is ejected. The aim of this work is to disentangle the inelastic scattering and {\small ICD} contributions to the formation of slow electrons.
      \label{fig:Ratio}}
  \end{center}
\end{figure}

\section{Disentanglement of ICD and inelastically scattered electrons}

To spectroscopically disentangle the {\small ICD} and inelastic scattering contributions to
electron-energy loss ({\small EEL}) electrons around $E_e = \SI{15}{eV}$
electron kinetic energy, the ratio of {\small EEL} to total electron counts is
divided by its off-resonant ($E_{h\nu} \le \SI{59.9}{eV}$) average which gives
the relative count ratio
\begin{equation}
  r(E_e) =
  \frac{c_{\text{\tiny EEL}}(E_e)}{c_{\text{total}}(E_e)}
  \left<\frac{c_{\text{\tiny EEL}}}{c_{\text{total}}}\right>_{\text{\!off}}^{-1}
\end{equation}
where $c$ denotes absolute counts and $\langle ~ \rangle_{\text{\!off}}$ the
average over the off-resonant photon energies.
If only inelastic scattering contributed to {\small EEL}, $c_{\text{\tiny EEL}}$
would be proportional to $c_{\text{total}}$ because the inelastic scattering
cross sections are nearly constant across the resonance. Consequently, the ratio
would be constant and $r(E_e) = 1$ across the tuning range of the photon energy studied here.
The {\small ICD} channel adds additional counts
$c_{\text{\tiny EEL}}(E_e)$ which results in values $r(E_e) > 1$ on the resonance.
The ratio of {\small ICD} to total {\small EEL} electrons is then
$\left(r - 1\right)/r$ which gives the number of {\small ICD} electrons
as a function of {\small EEL} or total electron counts
\begin{equation}
  c_{\text{\tiny ICD}}(E_e)
  = \frac{r(E_e) - 1}{r(E_e)} \, c_{\text{\tiny EEL}}(E_e)
  = \left(r(E_e) - 1\right) \,
  \left<\frac{c_{\text{\tiny EEL}}}{c_{\text{total}}}\right>_{\text{\!off}}
  \, c_{\text{total}}(E_e)
\end{equation}
and the probability of an electron to originate from the {\small ICD} process
\begin{equation}
  \label{eq:prob}
  p_{\text{\tiny ICD}}(E_e)
  = \frac{c_{\text{\tiny ICD}}(E_e)}{c_{\text{total}}(E_e)}
  = \frac{r(E_e) - 1}{r(E_e)} \, \frac{c_{\text{\tiny EEL}}(E_e)}{c_{\text{total}}(E_e)}
  = \left(r(E_e) - 1\right) \,
  \left<\frac{c_{\text{\tiny EEL}}}{c_{\text{total}}}\right>_{\text{\!off}} \;.
\end{equation}
Uncertainties are determined by Gaussian error propagation assuming $\sqrt{c}$
errors for the {\small EEL} and total electron counts and using the standard
deviation of the off-resonant $c_{\text{\tiny EEL}} / c_{\text{total}}$ values
for the error of their average. To evaluate the hemispherical electron analyzer
({\small HEA}) data for which the total electrons were not measured,
the {\small ICD} probability is factorized into
\begin{equation}
  \label{eq:factor}
  p_{\text{\tiny ICD}}(E_e)
  = \frac{c_{\text{\tiny ICD}}(E_e)}{c_{\text{total}}(E_e)}
  = \frac{c_{\text{\tiny ICD}}(E_e)}{c_{\text{\tiny EEL}}(E_e)}
    \frac{c_{\text{\tiny EEL}}(E_e)}{c_{\text{total}}(E_e)}
  = \frac{c_{\text{\tiny ICD}}(E_e)}{c_{\text{\tiny EEL}}(E_e)}
  \, p_{\text{\tiny EEL}}(E_e)
\end{equation}
and the second factor ($p_{\text{\tiny EEL}}$) is taken from {\small VMI} data with a similar average droplet
size.

\section{Baseline correction for count ratios from the HEA data}

\begin{figure}[ht]
  \begin{center}
    \includegraphics[width=.55\textwidth]{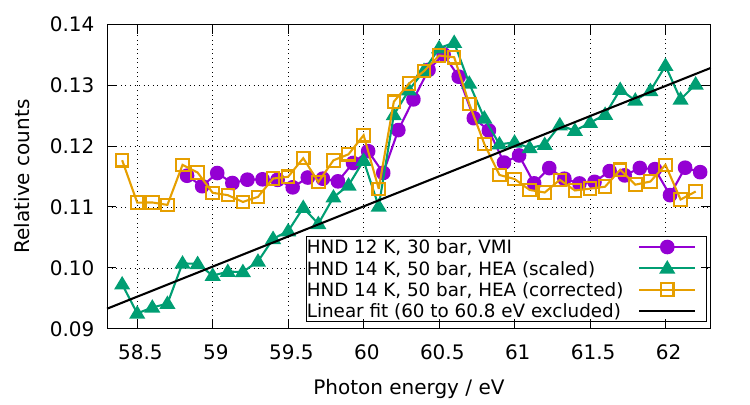}
    \caption{%
      Linear correction of the count ratio of low kinetic energy
      (\SIrange{18}{23}{eV}) to total electrons.
      {\small VMI} and {\small HEA} data are shown for the He nanodroplet
      ({\small HND}) expansion conditions specified in the legend.
      The {\small HEA} data are corrected by a linear fit to the
      off-resonant photon energy range around the double excitation resonance.
      The agreement of the corrected data with the result from velocity map
      imaging ({\small VMI}) data justifies this approach.
      \label{fig:linear-correction}}
  \end{center}
\end{figure}

\begin{figure}[ht]
    \hfill
    \begin{minipage}{.47\textwidth}
      \includegraphics[width=\textwidth]{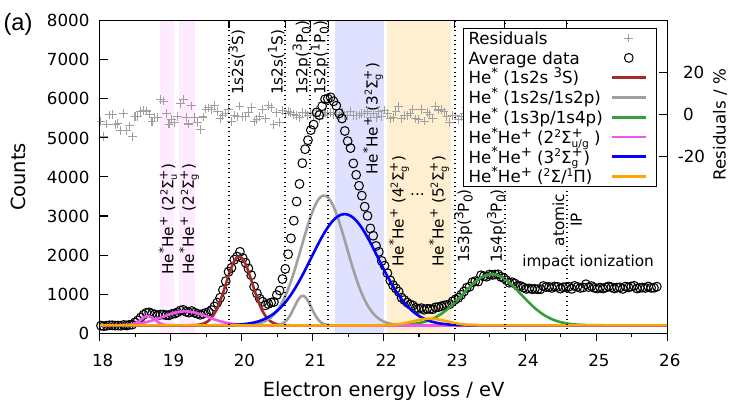}\\
      \includegraphics[width=\textwidth]{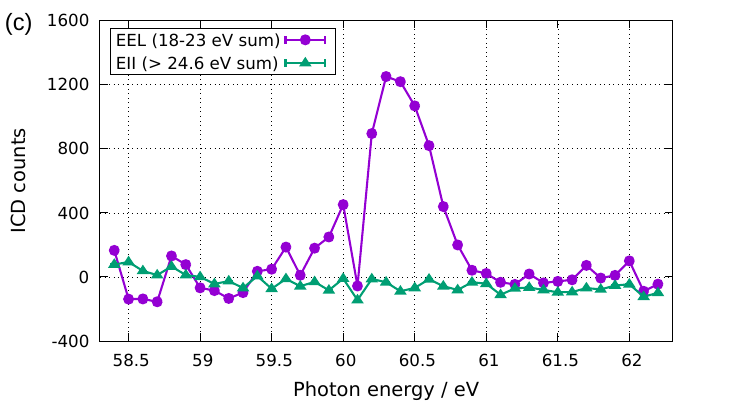}\\
      \includegraphics[width=\textwidth]{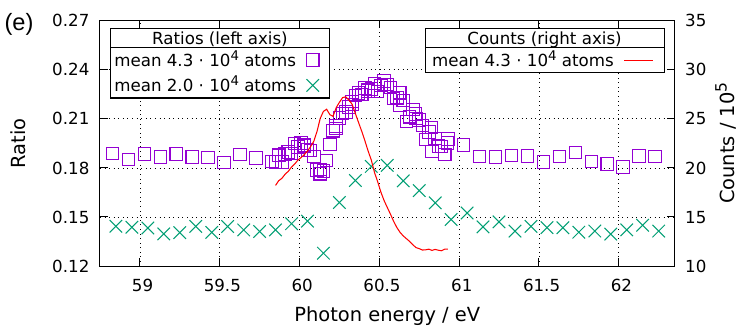}
    \end{minipage}
    ~~
    \begin{minipage}{.47\textwidth}
      \includegraphics[width=\textwidth]{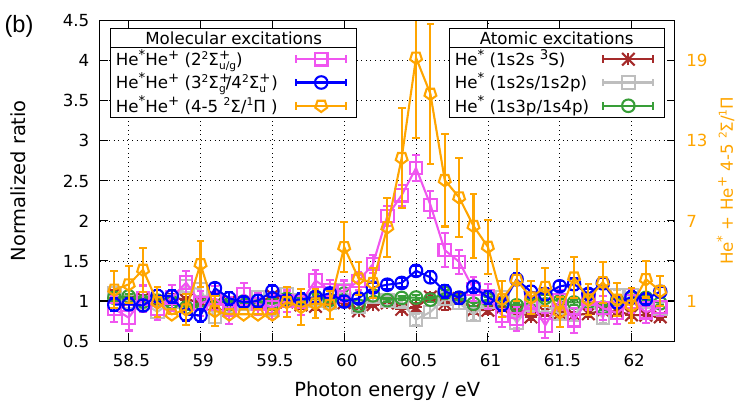}\\
      \includegraphics[width=\textwidth]{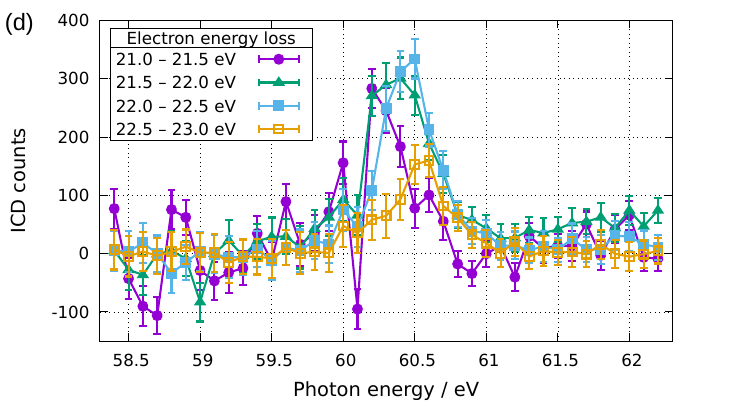}\\
      \includegraphics[width=\textwidth]{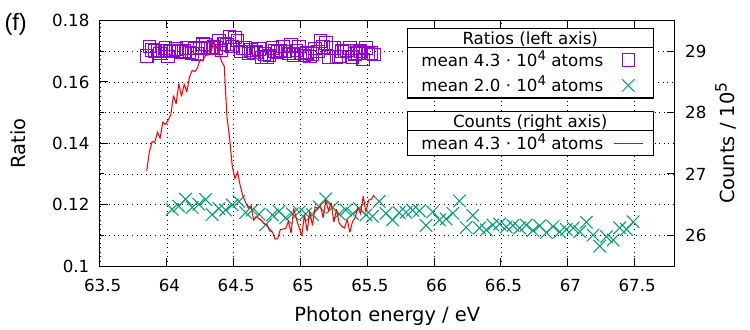}
    \end{minipage}\\
    \caption{%
      \textbf{(a)} Average {\small HEA} high-resolution {\small EEL} spectrum with fit residuals
      (gray crosses, right axis) and individual Gaussians (see labels in the legend)
      of the multi-peak fit that are not shown in Fig.~4~(a) in the main text.
      Black circles are averaged over 39 spectra recorded in the range $h\nu=58.4$\,--\,62.2~eV.
      Vertical dashed lines show atomic excitations with values from
      {\small NIST}~\cite{nist13}.
      Shaded areas correspond to He$^*$He$^+$ states with boundaries computed
      for {\small ICD} decay of the He$^{**}$ droplet excitation into the He$_2^{+*}$
      states shown in (c).
      \textbf{(b)} Photon energy dependence of the normalized peak area to total
      electron ratios for different states with colors corresponding to (a).
      In two cases (magenta and gray), the sum of two peak area is considered.
      Those components assigned to molecular excitations
      display pronounced maxima in the range of the Fano resonance.
      The ratios for the molecular 4--5~$^2\Sigma/^1\Pi$ excitations
      are given by the second $y$-axis on the right.
      \textbf{(c)} Estimate of the total {\small ICD} counts in the electron
      impact excitation ({\small EEL}) and electron impact ionization ({\small EII})
      range of the electron kinetic energy. Integration ranges of the electron
      energy loss are specified in the legend. The 1s3p/1s4p range is excluded
      because it overlaps with the {\small EII} region.
      \textbf{(d)} Photon energy dependence of {\small ICD} counts at different
      levels of molecular excitation in the 21\,--\,23~eV range. Higher
      excitation is preferentially observed at higher photon energies.
      \textbf{(e)} and \textbf{(f)}
      Ratios of {\small EEL} (18\,--\,23~eV e$^-$ kinetic energy loss) to
      photoelectron counts ($-$2\,--\,2~eV loss) as a function of photon
      energy at different {\small HND} sizes specified in the legend.
      Total photoelectrons at the $n=2$ and $n=3$ Fano resonances (assigned
      according to Ref.~\cite{laforge_fano_2016} Fig.~1) are shown as solid red
      lines (right axes). The lower photon energies~(e) feature the discussed
      {\small ICD} resonance around 60.5~eV photon energy. Assuming a similar
      {\small ICD} efficiency for the higher photon energies~(f) would give an
      increase of the order of 2\,\% which is in the noise level.
      \label{fig:EEL}}
\end{figure}

As photoelectrons were not detected in the {\small HEA} measurement, measured total
electron scans from a previous beamtime with the same droplet source but
a different detector were used~\cite{laforge_fano_2016}.
The resulting skewed baseline was corrected
with a constant slope which is justified by comparing to the present {\small VMI} data
where the ratios stem from a single measurement,
see Fig.~\ref{fig:linear-correction} and Fig.~2~(d) in the main text.
The derived normalized {\small EEL} to total electron ratios and total
{\small ICD} counts are presented Fig.~\ref{fig:EEL}~(b) and
Fig.~4~(b) in the main text, respectively.

\section{Count ratios from Fano and Lorentzian fits}

To evaluate the {\small ICD} probability in \eqref{eq:prob} or \eqref{eq:factor} and related
quantities, we fit the measured electron counts as a function of photon energy
with the Fano profile~\cite{hertel2008atoms}
\begin{equation}\label{eq:fano}
  c(h\nu)
  =
  s_{d}
  +
  s_{r} \, \frac{%
    \left(q + \frac{h\nu - E_{r}}{\Gamma}\right)^{2}
  }{%
    1 + \left(\frac{h\nu - E_{r}}{\Gamma}\right)^{2}
  }
\end{equation}
with photon energy $h\nu$, asymmetry (Fano) parameter $q$, resonance energy and
width $E_{r}$ and $\Gamma$, the resonant amplitude $s_{r}$ for double excitation
and the direct ionization amplitude $s_{d}$.
In this case, $q$ is negative and the maximum, resonance and minimum points are
\begin{align}
  \label{eq:points}
  &E_r + \Gamma/q,
  &
  &E_r
  &\text{and}&&
  E_r - \Gamma \, q
  \\
  \intertext{with the amplitudes}
  \label{eq:amplitudes}
  &s_d + \left(q^2 + 1\right) s_r,
  &
  &s_d + q^2 s_r
  &\text{and}&&
  s_d \,.
\end{align}
In case of {\small ICD} electrons only the resonant pathway is involved and the
shape is found to reduce to a symmetric Lorentzian profile with $q = 0$, $s_d + s_r
\approx 0$ and the maximum amplitude $s_d$ is found at the resonance energy.
To compute count ratios (relative probabilities), we evaluate all amplitudes at
the resonance energies obtained from the fit which gives $s_d + q^2 s_r$.
The resonance energy of all fits to the total, {\small EEL} and {\small ICD}
electrons is close to 60.4~eV.
All fit results are summarized in Table~\ref{tab:fits}.

\begin{table}
  \caption{\label{tab:fits}%
    Fit results from Fano and Lorentzian fits. The photon energy range from
    59.8~eV to 60.2~eV was excluded for the fits because of the sharply peaked
    photoline from an atomic He fraction in the droplet beam.
    The He$^+$ coincidences are most strongly affected so the fits for the
    total coincidence electrons are not reliable and impossible at 16~K.
    Counts ($s_d$, $s_r$ and $s_d + q^2 s_r$) are tabulated in units of $10^{6}$
    for all electrons (all $e^{-}$) and $10^{4}$ for He$_n^+$--$\,e^-$ coincidences
    and {\small HEA} data.
  }
  \footnotesize
  \begin{tabular}{cc@{~~}cccc@{~~}c@{~~}c@{~~}c@{~~}c@{~~}c}
    Det. & T/K & p/bar & Data & Electrons & $q$ & $E_r$/eV & $\Gamma$/eV & $s_d$ & $s_r$ & $s_d + q^2 s_r$ \\
    \hline 
    \input{table_1_a}%
    \input{table_1_b}%
    \input{table_1_c}%
    \input{table_1_d}%
    \input{table_1_e}  \end{tabular}
\end{table}

\section{Electron trapping probability}

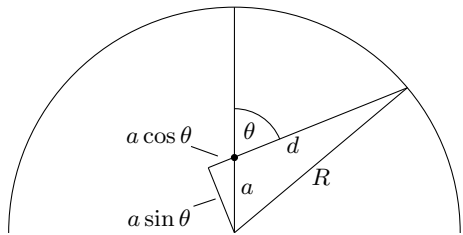
\begin{figure}[t]
  \begin{center}
    \usetikzlibrary {calc}
    \begin{tikzpicture}
      \path[clip] (-3.1,0) rectangle (3.1,3.1);
      \path (0,0) node (p0) {};
      \draw (0,0) circle(3);
      \draw (0,0) -- (0,3);
      \draw (0,0) -- node[below] {$R$} (40:3) node(p1) {};
      \path (40:3) -- (-2.5,0) node(p2) {};
      \draw (0,0) -- ($(p1)!(p0)!(p2)$);
      \draw ($(p1)!(p0)!(p2)$) -- node[below left] {$d$} (40:3);
      \node at (0.17,0.58) {$a$};
      \node[inner sep=1pt](sin) at (-1,0.2) {$a \sin\theta$};
      \draw (sin) -- (-0.25,0.45);
      \node[inner sep=1pt](cos) at (-1,1.3) {$a \cos\theta$};
      \draw (cos) -- (-0.22,1.02);
      \node at (0.2,1.35) {$\theta$} -- (0,1);
      \draw[fill] (0,1) circle[radius=0.04];
      \draw (0,1.65) arc (90:22:0.65);
    \end{tikzpicture}
    \caption{%
      Geometrical construction of the distance $d$ of a point in a sphere from a
      point on its surface. The distance of the inner point from the center of
      the sphere with radius $R$ is denoted by $a$.
      One finds \(
        \left(d + a\cos\left(\theta\right)\right)^2
        = R^2 - a^2\sin^2\left(\theta\right)
      \).
      \label{fig:geometry}}
  \end{center}
\end{figure}

Trapping of slow {\small ICD} electrons in helium droplets with increasing size
affects the probability of their observation~\cite{asmussen2023dopant,asmussen2023electron}.
It is modelled as a product of the size independent {\small ICD} probability
$p_0$ and the likelihood for electron ejection before trapping (see fit in Fig.~3 in the main text),
\begin{align}\label{eq:pICD}
  p^{\rm obs}_{\rm ICD} &= p_0 \exp\left(-\frac{\bar d}{l_{\rm slow}}\right)
  = p_0 \exp\left(- \frac{3 R}{4 l_{\rm slow}}\right)
\end{align}
where $\bar d = \frac{3}{4} R$ is the average travelling distance to the surface
in a droplet with radius $R$ and $l_{\rm slow}$ is the mean free path of the electron.
We have thus neglected the angular deflection by elastic and inelastic
collisions that slow down the electron and roughly estimate the ejection
probability alone from the average distance $\bar d$ from the surface.
The droplet radius is computed from the number of atoms $N_{\rm He}$ in
the droplet by the expression
\begin{align}\label{eq:size}
  R / \text{nm} &= 0.22 \sqrt[3]{N_{\rm He}}
\end{align}
by Gomez~\textit{et al.}~\cite{Gomez2011}.
The expression for the mean distance $\bar d$ is derived as follows.
From Fig.~\ref{fig:geometry} we find the distance $d$ of an electron
from the surface of a spherical droplet with radius $R$,
\begin{align}
  d &= \sqrt{R^2 - a^2 \sin^2(\theta)} - a \cos(\theta)
\end{align}
where $a$ is the distance from the electron to the center of the droplet.
The isotropic likelihood that the electron originates from a distance $a$
from the center and moves toward a surface element
$[\theta, \theta + {\rm d}\theta] \times [\phi, \phi + {\rm d}\phi]$
is given by
\begin{align}
  \frac{3 a^2}{R^3} \,{\rm d}a ~\times~
  \frac{1}{4 \pi} \sin(\theta) \,{\rm d}\theta \,{\rm d}\phi
\end{align}
and we obtain the distance averaged over all positions and directions
by integration,
\begin{align}
  \bar d &= \frac{3}{4 \pi R^3} \int_{0}^{R} \int_{0}^{\pi} \int_{0}^{2\pi}
  a^2 \left( \sqrt{R^2 - a^2 \sin^2(\theta)} - a \cos(\theta) \right)
  \sin(\theta) \,{\rm d}\phi \,{\rm d}\theta \,{\rm d} a
  \\\nonumber
  &= \frac{3}{2 R^3} \int_{0}^{R} \int_{0}^{\pi}
  \left( \sqrt{R^2 - a^2 \sin^2(\theta)} - a \cos(\theta) \right)
  a^2 \sin(\theta) \,{\rm d}\theta \,{\rm d} a
  \quad \text{(the $\cos(\theta) \sin(\theta)$ term vanishes by symmetry)}
  \\\nonumber
  &= \frac{3 R}{2} \int_{0}^{1} \int_{0}^{\pi}
  \sqrt{1 - k^2 \sin^2(\theta)}
  \; k^2 \sin(\theta) \,{\rm d}\theta \,{\rm d} k
  = \frac{3 R}{2} \int_{0}^{1} \int_{-1}^{1}
  \sqrt{1 - k^2 \left(1 - \mu^2\right)}
  \; k^2 \,{\rm d}\mu \,{\rm d} k
  \\\nonumber
  &= \frac{3 R}{2} \int_{0}^{1} k \int_{-k}^{k}
  \sqrt{x^2 + 1 - k^2} \; \,{\rm d} x \,{\rm d} k
  = \frac{3 R}{2} \int_{0}^{1} \frac{k}{2}
  \left[x \sqrt{x^2 + 1 - k^2} + (1 - k^2)
  \ln\left(x + \sqrt{x^2 + 1 - k^2}\right)\right]_{-k}^{k} \,{\rm d} k
  \\\nonumber
  &= \frac{3 R}{2} \int_{0}^{1} \frac{k}{2}
  \left(k - (-k) + (1 - k^2)
  \ln\left(\frac{1 + k}{1 - k}\right)\right) \,{\rm d} k
  = \frac{3 R}{4} \int_{0}^{1} \left(
    2 k^2 + \left(k - k^3\right) \ln\left(\frac{1 + k}{1 - k}\right)
  \right) \,{\rm d} k
  = \frac{3 R}{4} \;.
\end{align}

\section{Electron impact excitation probability}

\begin{figure}[t]
  \begin{center}
    \includegraphics[width=0.8\textwidth]{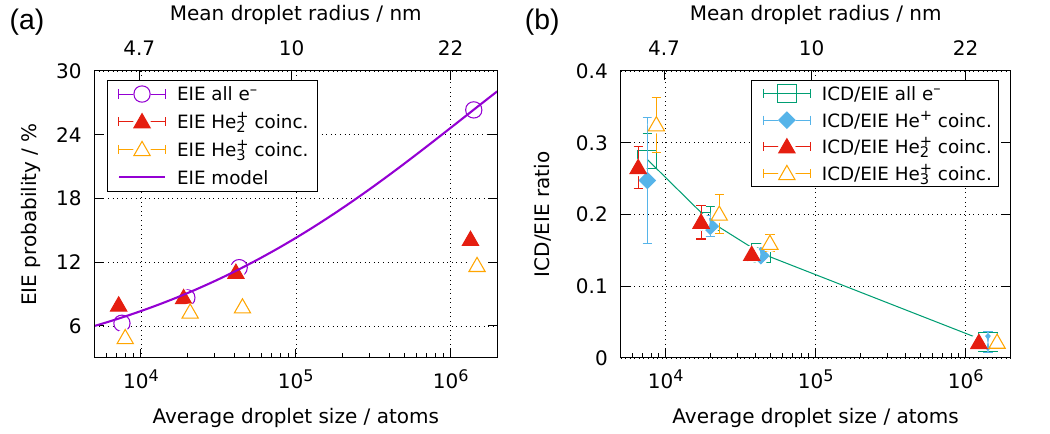}
    \begin{minipage}{0.9\textwidth}
    \caption{%
      Droplet-size dependence for total electrons and coincidences with
      He$_{n}^{+}$ cations, respectively. The points for $n=2,3$ are somewhat
      shifted to lower and higher droplet sizes for better visibility.
      \textbf{(a)} {\small EIE} probability relative to total electrons
      (all e$^{-}$) or coincidences (He$_{n}^{+}$, $n=2,3$) with the
      {\small EIE} model fitted to the total electron data.
      \textbf{(b)} Relative ratio of {\small ICD} and {\small EIE} (electron
      impact excitation of 1s2s/1s2p states) counts. The reen solid line
      connects the total electron points to guide the eye.
      \label{fig:size}}
    \end{minipage}
  \end{center}
\end{figure}

The loss of fast electrons is predominated by electron impact excitation
({\small EIE}) and ionization ({\small EII}) that we describe in a simple model
by the average travelling distance $\bar d$ to the surface that has been derived
in the previous section and a mean free path $l_{\rm fast}$ in analogy to
Eq.~\eqref{eq:pICD}.
The {\small EIE} probability is then given by
\begin{align}\label{eq:pEEL}
  P_{\rm EIE} / \text{\small \%}
  &= p_{\rm EIE} \left(1 - \exp\left(-\frac{\bar d}{l_{\rm fast}}\right)\right)
   = p_{\rm EIE} \left(1 - \exp\left(- \frac{3 R}{4 l_{\rm fast}}\right)\right)
\end{align}
where $p_{\rm EIE}$ is the fraction of {\small EIE} from the total
inelastic scattering cross section ({\small EIE} = E{\small EL} $-$ {\small ICD}).
The fit in Fig.~\ref{fig:size}~(a) gives the mean free path for inelastic
scattering $l_{\rm fast} = (17.6 \pm 1.4)$~nm and a $(40 \pm 2)$\,\% fraction
of {\small EIE} of the 1s2s/1s2p states relative to the total inelastic cross section in
close agreement with \SI{17}{nm} and \SI{38}{\percent} in Sec.~\ref{sec:eie}
derived from the cross sections in Ref.~\cite{ralchenko2008electron}.
Surprisingly, the loss of slow electrons thus seems to be negligible in the
investigated droplet size range, in opposition to the size dependence of the
ICD probability that indicates suppressed ejection of slow electrons.
This discrepancy is only partially explained by the fact that fast electrons
typically travel a substantial distance to the droplet surface before losing
kinetic energy by inelastic scattering. A consistent explanation of the observed
size dependences would thus require an improved model considering the different
pathways for electrons propagating in a droplet instead of using a single
average distance from the surface.

The ratios of {\small EIE} coincidences to total coincidences in
Fig.~\ref{fig:size}~(a) do increase with growing {\small EEL} probability at larger
droplet sizes, but the growth is significantly weaker than for all electrons
because the cations are more easily trapped in the droplets than
electrons~\cite{asmussen2023dopant}.
At all sizes, the {\small ICD}/{\small EIE} ratio
in Fig.~\ref{fig:size}~(b) is within uncertainties
identical for total electrons and e$^{-}$--He$_{n}^{+}$, $n=1,2,3$,
coincidences. There is
consequently no evidence that the {\small ICD} process at short distances in the
He$^*$He$^+$ potential energy curves affects the ion formation and ejection
dynamics.

\section{\label{sec:eie}Electron impact cross section}

To estimate the mean free path $l_{\rm fast}$ for inelastic scattering, we sum up the
electron impact ionization and excitation cross sections from the initial
1\,\textsuperscript{1}S state at $E_e = \SI{35.8}{eV}$ electron energy
(corresponding to $E_{h\nu} = \SI{60.4}{eV}$ photon energy) according to graphs
4\,--\,6 in Ref.~\cite{ralchenko2008electron} which gives the cross section
$\sigma_{\rm total} \approx \SI{0.27}{\AA^2}$.
With the helium density
$n_{\rm He} = N_{\rm He}/(4/3\pi R^3) \approx \SI{22.42}{nm^{-3}}$ from
Eq.~\eqref{eq:size} we obtain the corresponding mean free path
\begin{align}
  l_{\rm fast} = \frac{1}{\sigma_{\rm exc} \, n_{\rm He}} \approx \SI{17}{nm}.
\end{align}
In the main text, {\small EEL} denotes all electrons with 18 to 23~eV energy
loss, excluding excitation of 1s3p and 1s4p states that overlap with electron
impact ionization.
Subtracting the {\small ICD} contribution gives the {\small EIE} yield that
consequently only includes excitation of 1s2s and 1s2p states with the
cross section $\sigma_{\rm 1s2s/1s2p} \approx \SI{0.10}{\AA^2}$
in Ref.~\cite{ralchenko2008electron}.
Its ratio to the total inelastic cross section is
\begin{align}
  p_{\rm EIE} = \frac{\sigma_{\rm 1s2s/1s2p}}{\sigma_{\rm total}} \approx 0.38 \,.
\end{align}

\section{Theoretical estimate of the ICD probability}

\begin{figure}[t]
  \begin{center}
    \includegraphics[width=0.66\textwidth]{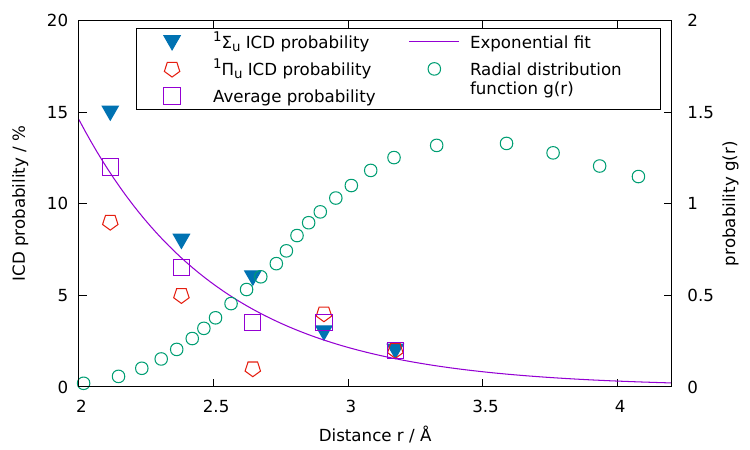}
    \begin{minipage}{0.9\textwidth}
    \caption{%
      Computed {\small ICD} probabilities for the He(2s2p) -- He(1s$^2$)
      $^1\Sigma_u$ and $^1\Pi_u$ states with an exponential fit to their
      average (magenta squares). Green circles represent the pair-distance distribution
      function $g(r)$~\cite{Chin1995:prb,peterka2007photoionization}.
      \label{fig:icd_theory}}
    \end{minipage}
  \end{center}
\end{figure}

Total and partial decay widths of the He(2s2p)--He(1s$^2$) $^1\Sigma_u$ and
$^1\Pi_u$ states have been computed as described in the main text at five
different distances. The resulting autoionization ({\small AI}) probabilities
are summarized in Table~\ref{tab:decay_widths}. The differences of 100\,\% and
the {\small AI} probabilities are the total probabilities for the decay into
different {\small ICD} channels as a function of distance which is plotted in
Fig.~\ref{fig:icd_theory}. We are not aware of available data on the relative
excitation of the $^1\Sigma_u$ and $^1\Pi_u$ states and the respective
{\small ICD} probabilities are not differing much. Particularly at larger
distances, they are close to zero such that the computational method can only
give an estimate with limited accuracy. Therefore, we average the probabilities
at each distance with equal weight and the underlying assumption of similar
cross sections has only a small effect on the resulting average.
\begin{table}[ht]
  \caption{\label{tab:decay_widths}%
    Likelihood of {\small AI} for $^1\Sigma_u$ and $^1\Pi_u$
    states at different pair distances $r$.
  }
  \begin{tabular}{l@{\quad}l@{\quad}l@{\quad}l@{\quad}l@{\quad}l}
    r (a.\,u.) & 6    & 5.5  & 5    & 4.5  & 4    \\
    r (\AA)    & 3.18 & 2.91 & 2.65 & 2.38 & 2.12 \\
    \hline
    He(2s2p) -- He(1s$^2$) $^1\Sigma_u$ $\rightarrow$ AI (\%)
                & 98   & 97   & 94   & 92   & 85 \\
    He(2s2p) -- He(1s$^2$) $^1\Pi_u$ $\rightarrow$ AI (\%)
                & 98   & 96   & 99   & 95   & 91 \\
  \end{tabular}
\end{table}

To derive the overall {\small ICD} probability,
the distance dependent average probabilities are fitted with an exponential
curve. This curve is weighted with the pair-distance distribution function
$g(r)$~\cite{Chin1995:prb,peterka2007photoionization} to obtain the total
{\small ICD} probability relative to the
resonant double excitation cross section ({\small AI} plus {\small ICD})
which is $(3.0 \pm 2.8)\,\%$. The large uncertainty is due to the
standard deviations of the fit parameters $a = 670 \pm 380$,
$b = 1.9 \pm 0.3$, $p(r) = a \exp\left(-b r\right)$.
The {\small ICD} probability relative to the total ionization cross section
is obtained by multiplying this result with the ratio $f_\text{res}$ of
the resonant cross section to the total electron counts $c(h\nu)$
in Eq.~\eqref{eq:fano} evaluated at the
double excitation resonance~($h\nu = E_{r}$),
\begin{equation}\label{eq:fres}
  f_\text{res}
  = \frac{s_r q^2}{s_d + s_r q^2} \left(1 + \frac{2}{\pi q^2}\right)^{-1}
\end{equation}
which is a product of two ratios. According to Ref.~\cite{Mezei2015:epjw},
we identify a constant background contribution $s_{d}$ and the unperturbed
continuum cross section $s_{r}$ in the Fano profile in Eq.~\eqref{eq:fano}.
Then the total cross section $c(h\nu)$ is a sum of the background $s_{d}$ and
a transition probability\,---\,which we denote as $c_r(h\nu)$\,---\,that
stems from the Fano interference of the continuum with the discrete level
(double excitation leading to {\small AI} or {\small ICD}).
At the resonance, the latter simplifies to $s_{r} q^{2}$ and its ratio to
the total cross section, $c_r(E_r)/c(E_r)$, is the first factor in Eq.~\eqref{eq:fres}.
For the second factor, we use the identification of $\frac{1}{2} \pi q^2$ in
Ref.~\cite{fano_effects_1961} as the ratio of the transition probabilities
to the modified discrete level ($A$) and to a band width of unperturbed
continuum states ($B$).  Assuming that the sum $A + B$ approximates the
transition probability $c_r(h\nu)$, its fraction of the resonant pathway is
\begin{equation}
  \frac{A}{c_r(h\nu)} \approx \frac{A}{A + B} = \frac{A/B}{A/B + 1}
  = \frac{\frac{1}{2} \pi q^2}{\frac{1}{2} \pi q^2 + 1}
  = \frac{1}{1 + \frac{2}{\pi q^2}}
\end{equation}
which is the second factor needed to compute $f_\text{res}$.
With the fit results for total electrons in Table~\ref{tab:fits} we obtain the
ratios in Table~\ref{tab:res_fraction} for four different nozzle temperatures.
The values do not significantly differ from each other and we use the average
$f_\text{res} = 0.51 \pm 0.03$ as final result.

\begin{table}[ht]
  \caption{\label{tab:res_fraction}%
  Fractions of resonant to total ionization according to Eq.~\ref{eq:fres}
  for {\small VMI} data at different nozzle temperatures.
  }
  \begin{tabular}{c@{~~}|@{~~}c@{~~}@{~~}c@{~~}@{~~}c@{~~}@{~~}c}
    Nozzle temperature & 10 K & 12 K & 14 K & 16 K \\
    \hline
    $f_\text{res}$
    & $0.57 \pm 0.09$
    & $0.47 \pm 0.02$
    & $0.50 \pm 0.04$
    & $0.49 \pm 0.04$ \\
  \end{tabular}
\end{table}

\FloatBarrier
\bibliography{Bib}

%% file: table_1_a.tex
    % electron_eKE_loss/fano_fits_total_e/fit_Astrid_10_K_data.txt (total)
    {\scriptsize VMI} & 10 & 30 & all $e^{-}$ & total
                 & $-8.21 \pm 1.21$ % q
                 & $60.349 \pm 0.006$ % E_r
                 & $0.217 \pm 0.005$ % Gamma
                 & $0.800 \pm 0.015$ % s_d
                 & $0.016 \pm 0.004$ % s_r
                 & $1.9 \pm 0.4$ % s_d + q^2 s_r
                 \\
    % electron_eKE_loss/fano_fits_total_e/fit_Astrid_10_K_data.txt (EEL)
    {\scriptsize VMI} & 10 & 30 & all $e^{-}$ & {\scriptsize EEL}
                 & $-8.63 \pm 1.32$ % q
                 & $60.354 \pm 0.006$ % E_r
                 & $0.220 \pm 0.005$ % Gamma
                 & $0.201 \pm 0.004$ % s_d
                 & $0.004 \pm 0.001$ % s_r
                 & $0.50 \pm 0.12$ % s_d + q^2 s_r
                 \\
    % electron_eKE_loss/fano_fits_total_e/fit_Astrid_10_K_data.txt (ICD)
    {\scriptsize VMI} & 10 & 30 & all $e^{-}$ & {\scriptsize ICD}
                 & ~~$0$ % q
                 & $60.415 \pm 0.003$ % E_r
                 & $0.220 \pm 0.009$ % Gamma
                 & $0.011 \pm 0.000$ % s_d
                 & $-0.022 \pm 0.001$ % s_r
                 & $0.0110 \pm 0.0003$ % s_d + q^2 s_r
                 \\
    % electron_eKE_loss/fano_fits_total_e/fit_Astrid_12_K_data.txt (total)
    {\scriptsize VMI} & 12 & 30 & all $e^{-}$ & total
                 & $-2.36 \pm 0.07$ % q
                 & $60.402 \pm 0.005$ % E_r
                 & $0.233 \pm 0.005$ % Gamma
                 & $1.361 \pm 0.011$ % s_d
                 & $0.272 \pm 0.015$ % s_r
                 & $2.88 \pm 0.12$ % s_d + q^2 s_r
                 \\
    % electron_eKE_loss/fano_fits_total_e/fit_Astrid_12_K_data.txt (EEL)
    {\scriptsize VMI} & 12 & 30 & all $e^{-}$ & {\scriptsize EEL}
                 & $-3.10 \pm 0.14$ % q
                 & $60.409 \pm 0.006$ % E_r
                 & $0.220 \pm 0.006$ % Gamma
                 & $0.161 \pm 0.002$ % s_d
                 & $0.023 \pm 0.002$ % s_r
                 & $0.38 \pm 0.03$ % s_d + q^2 s_r
                 \\
    % electron_eKE_loss/fano_fits_total_e/fit_Astrid_12_K_data.txt (ICD)
    {\scriptsize VMI} & 12 & 30 & all $e^{-}$ & {\scriptsize ICD}
                 & ~~$0$ % q
                 & $60.428 \pm 0.006$ % E_r
                 & $0.183 \pm 0.012$ % Gamma
                 & $0.056 \pm 0.002$ % s_d
                 & $-0.059 \pm 0.002$ % s_r
                 & $0.056 \pm 0.002$ % s_d + q^2 s_r
                 \\
    % electron_eKE_loss/fano_fits_total_e/fit_Astrid_14_K_data.txt (total)
    {\scriptsize VMI} & 14 & 30 & all $e^{-}$ & total
                 & $-3.03 \pm 0.18$ % q
                 & $60.365 \pm 0.009$ % E_r
                 & $0.245 \pm 0.010$ % Gamma
                 & $4.441 \pm 0.060$ % s_d
                 & $0.567 \pm 0.064$ % s_r
                 & $9.6 \pm 0.9$ % s_d + q^2 s_r
                 \\
    % electron_eKE_loss/fano_fits_total_e/fit_Astrid_14_K_data.txt (EEL)
    {\scriptsize VMI} & 14 & 30 & all $e^{-}$ & {\scriptsize EEL}
                 & $-3.91 \pm 0.27$ % q
                 & $60.401 \pm 0.007$ % E_r
                 & $0.219 \pm 0.008$ % Gamma
                 & $0.413 \pm 0.006$ % s_d
                 & $0.040 \pm 0.005$ % s_r
                 & $1.02 \pm 0.11$ % s_d + q^2 s_r
                 \\
    % electron_eKE_loss/fano_fits_total_e/fit_Astrid_14_K_data.txt (ICD)
    {\scriptsize VMI} & 14 & 30 & all $e^{-}$ & {\scriptsize ICD}
                 & ~~$0$ % q
                 & $60.439 \pm 0.010$ % E_r
                 & $0.173 \pm 0.019$ % Gamma
                 & $0.196 \pm 0.013$ % s_d
                 & $-0.210 \pm 0.013$ % s_r
                 & $0.196 \pm 0.013$ % s_d + q^2 s_r
                 \\
    % electron_eKE_loss/fano_fits_total_e/fit_Astrid_16_K_data.txt (total)
    {\scriptsize VMI} & 16 & 30 & all $e^{-}$ & total
                 & $-2.34 \pm 0.13$ % q
                 & $60.391 \pm 0.014$ % E_r
                 & $0.276 \pm 0.013$ % Gamma
                 & $0.522 \pm 0.008$ % s_d
                 & $0.121 \pm 0.011$ % s_r
                 & $1.18 \pm 0.09$ % s_d + q^2 s_r
                 \\
    % electron_eKE_loss/fano_fits_total_e/fit_Astrid_16_K_data.txt (EEL)
    {\scriptsize VMI} & 16 & 30 & all $e^{-}$ & {\scriptsize EEL}
                 & $-2.41 \pm 0.12$ % q
                 & $60.491 \pm 0.011$ % E_r
                 & $0.243 \pm 0.012$ % Gamma
                 & $0.040 \pm 0.001$ % s_d
                 & $0.011 \pm 0.001$ % s_r
                 & $0.104 \pm 0.009$ % s_d + q^2 s_r
                 \\
    % electron_eKE_loss/fano_fits_total_e/fit_Astrid_16_K_data.txt (ICD)
    {\scriptsize VMI} & 16 & 30 & all $e^{-}$ & {\scriptsize ICD}
                 & ~~$0$ % q
                 & $60.483 \pm 0.014$ % E_r
                 & $0.147 \pm 0.025$ % Gamma
                 & $0.028 \pm 0.003$ % s_d
                 & $-0.030 \pm 0.003$ % s_r
                 & $0.028 \pm 0.003$ % s_d + q^2 s_r
                 \\

%% file: table_1_b.tex
    % electron_eKE_loss/fano_fits_he_coinc/fit_Astrid_10_K_data.txt (total)
    {\scriptsize VMI} & 10 & 30 & He$^{+}$ coinc. & total
                 & $-3.39 \pm 1.96$ % q
                 & $60.300 \pm 0.043$ % E_r
                 & $0.344 \pm 0.033$ % Gamma
                 & $0.118 \pm 0.005$ % s_d
                 & $0.007 \pm 0.007$ % s_r
                 & $0.19 \pm 0.12$ % s_d + q^2 s_r
                 \\
    % electron_eKE_loss/fano_fits_he_coinc/fit_Astrid_10_K_data.txt (EEL)
    {\scriptsize VMI} & 10 & 30 & He$^{+}$ coinc. & {\scriptsize EEL}
                 & $-24.56 \pm 59.89$ % q
                 & $60.335 \pm 0.032$ % E_r
                 & $0.263 \pm 0.020$ % Gamma
                 & $0.032 \pm 0.004$ % s_d
                 & $0.000 \pm 0.000$ % s_r
                 & $0.032 \pm 0.004$ % s_d + q^2 s_r
                 \\
    % electron_eKE_loss/fano_fits_he_coinc/fit_Astrid_10_K_data.txt (ICD)
    {\scriptsize VMI} & 10 & 30 & He$^{+}$ coinc. & {\scriptsize ICD}
                 & ~~$0$ % q
                 & $60.418 \pm 0.002$ % E_r
                 & $0.235 \pm 0.008$ % Gamma
                 & $0.002 \pm 0.000$ % s_d
                 & $-0.004 \pm 0.000$ % s_r
                 & $0.00182 \pm 0.00004$ % s_d + q^2 s_r
                 \\
    % electron_eKE_loss/fano_fits_he_coinc/fit_Astrid_12_K_data.txt (total)
    {\scriptsize VMI} & 12 & 30 & He$^{+}$ coinc. & total
                 & $-2.04 \pm 0.12$ % q
                 & $60.435 \pm 0.013$ % E_r
                 & $0.277 \pm 0.015$ % Gamma
                 & $1.130 \pm 0.013$ % s_d
                 & $0.174 \pm 0.018$ % s_r
                 & $1.85 \pm 0.11$ % s_d + q^2 s_r
                 \\
    % electron_eKE_loss/fano_fits_he_coinc/fit_Astrid_12_K_data.txt (EEL)
    {\scriptsize VMI} & 12 & 30 & He$^{+}$ coinc. & {\scriptsize EEL}
                 & $-4.07 \pm 0.32$ % q
                 & $60.398 \pm 0.009$ % E_r
                 & $0.255 \pm 0.012$ % Gamma
                 & $0.277 \pm 0.006$ % s_d
                 & $0.029 \pm 0.004$ % s_r
                 & $0.76 \pm 0.10$ % s_d + q^2 s_r
                 \\
    % electron_eKE_loss/fano_fits_he_coinc/fit_Astrid_12_K_data.txt (ICD)
    {\scriptsize VMI} & 12 & 30 & He$^{+}$ coinc. & {\scriptsize ICD}
                 & ~~$0$ % q
                 & $60.420 \pm 0.004$ % E_r
                 & $0.205 \pm 0.009$ % Gamma
                 & $0.108 \pm 0.002$ % s_d
                 & $-0.113 \pm 0.003$ % s_r
                 & $0.108 \pm 0.002$ % s_d + q^2 s_r
                 \\
    % electron_eKE_loss/fano_fits_he_coinc/fit_Astrid_14_K_data.txt (total)
    {\scriptsize VMI} & 14 & 30 & He$^{+}$ coinc. & total
                 & $-2.12 \pm 0.33$ % q
                 & $60.457 \pm 0.035$ % E_r
                 & $0.300 \pm 0.038$ % Gamma
                 & $2.001 \pm 0.019$ % s_d
                 & $0.097 \pm 0.025$ % s_r
                 & $2.44 \pm 0.18$ % s_d + q^2 s_r
                 \\
    % electron_eKE_loss/fano_fits_he_coinc/fit_Astrid_14_K_data.txt (EEL)
    {\scriptsize VMI} & 14 & 30 & He$^{+}$ coinc. & {\scriptsize EEL}
                 & $-6.04 \pm 1.09$ % q
                 & $60.376 \pm 0.014$ % E_r
                 & $0.236 \pm 0.018$ % Gamma
                 & $0.482 \pm 0.013$ % s_d
                 & $0.018 \pm 0.006$ % s_r
                 & $1.1 \pm 0.3$ % s_d + q^2 s_r
                 \\
    % electron_eKE_loss/fano_fits_he_coinc/fit_Astrid_14_K_data.txt (ICD)
    {\scriptsize VMI} & 14 & 30 & He$^{+}$ coinc. & {\scriptsize ICD}
                 & ~~$0$ % q
                 & $60.429 \pm 0.006$ % E_r
                 & $0.204 \pm 0.013$ % Gamma
                 & $0.207 \pm 0.007$ % s_d
                 & $-0.219 \pm 0.007$ % s_r
                 & $0.207 \pm 0.007$ % s_d + q^2 s_r
                 \\
    % electron_eKE_loss/fano_fits_he_coinc/fit_Astrid_16_K_data.txt (total)
    {\scriptsize VMI} & 16 & 30 & He$^{+}$ coinc. & total
                 & $-3.41 \pm 0.47$ % q
                 & $60.154 \pm 0.010$ % E_r
                 & $0.040 \pm 0.006$ % Gamma
                 & $1.021 \pm 0.097$ % s_d
                 & $0.298 \pm 0.104$ % s_r
                 & $4.5 \pm 1.5$ % s_d + q^2 s_r
                 \\
    % electron_eKE_loss/fano_fits_he_coinc/fit_Astrid_16_K_data.txt (EEL)
    {\scriptsize VMI} & 16 & 30 & He$^{+}$ coinc. & {\scriptsize EEL}
                 & $-3.51 \pm 0.30$ % q
                 & $60.453 \pm 0.019$ % E_r
                 & $0.289 \pm 0.021$ % Gamma
                 & $0.123 \pm 0.003$ % s_d
                 & $0.017 \pm 0.003$ % s_r
                 & $0.33 \pm 0.05$ % s_d + q^2 s_r
                 \\
    % electron_eKE_loss/fano_fits_he_coinc/fit_Astrid_16_K_data.txt (ICD)
    {\scriptsize VMI} & 16 & 30 & He$^{+}$ coinc. & {\scriptsize ICD}
                 & ~~$0$ % q
                 & $60.459 \pm 0.013$ % E_r
                 & $0.200 \pm 0.026$ % Gamma
                 & $0.084 \pm 0.005$ % s_d
                 & $-0.089 \pm 0.005$ % s_r
                 & $0.084 \pm 0.005$ % s_d + q^2 s_r
                 \\

%% file: table_1_c.tex
    % electron_eKE_loss/fano_fits_he2_coinc/fit_Astrid_10_K_data.txt (total)
    {\scriptsize VMI} & 10 & 30 & He$_{2}^{+}$ coinc. & total
                 & $-10.11 \pm 1.70$ % q
                 & $60.299 \pm 0.005$ % E_r
                 & $0.242 \pm 0.004$ % Gamma
                 & $0.380 \pm 0.005$ % s_d
                 & $0.004 \pm 0.001$ % s_r
                 & $0.79 \pm 0.17$ % s_d + q^2 s_r
                 \\
    % electron_eKE_loss/fano_fits_he2_coinc/fit_Astrid_10_K_data.txt (EEL)
    {\scriptsize VMI} & 10 & 30 & He$_{2}^{+}$ coinc. & {\scriptsize EEL}
                 & $-10.76 \pm 6.73$ % q
                 & $60.312 \pm 0.020$ % E_r
                 & $0.258 \pm 0.015$ % Gamma
                 & $0.054 \pm 0.003$ % s_d
                 & $0.001 \pm 0.001$ % s_r
                 & $0.17 \pm 0.19$ % s_d + q^2 s_r
                 \\
    % electron_eKE_loss/fano_fits_he2_coinc/fit_Astrid_10_K_data.txt (ICD)
    {\scriptsize VMI} & 10 & 30 & He$_{2}^{+}$ coinc. & {\scriptsize ICD}
                 & ~~$0$ % q
                 & $60.417 \pm 0.003$ % E_r
                 & $0.247 \pm 0.012$ % Gamma
                 & $0.003 \pm 0.000$ % s_d
                 & $-0.006 \pm 0.000$ % s_r
                 & $0.0025 \pm 0.0001$ % s_d + q^2 s_r
                 \\
    % electron_eKE_loss/fano_fits_he2_coinc/fit_Astrid_12_K_data.txt (total)
    {\scriptsize VMI} & 12 & 30 & He$_{2}^{+}$ coinc. & total
                 & $-2.39 \pm 0.10$ % q
                 & $60.384 \pm 0.007$ % E_r
                 & $0.232 \pm 0.007$ % Gamma
                 & $4.210 \pm 0.042$ % s_d
                 & $0.757 \pm 0.056$ % s_r
                 & $8.5 \pm 0.5$ % s_d + q^2 s_r
                 \\
    % electron_eKE_loss/fano_fits_he2_coinc/fit_Astrid_12_K_data.txt (EEL)
    {\scriptsize VMI} & 12 & 30 & He$_{2}^{+}$ coinc. & {\scriptsize EEL}
                 & $-3.19 \pm 0.21$ % q
                 & $60.400 \pm 0.009$ % E_r
                 & $0.240 \pm 0.009$ % Gamma
                 & $0.486 \pm 0.007$ % s_d
                 & $0.060 \pm 0.007$ % s_r
                 & $1.10 \pm 0.11$ % s_d + q^2 s_r
                 \\
    % electron_eKE_loss/fano_fits_he2_coinc/fit_Astrid_12_K_data.txt (ICD)
    {\scriptsize VMI} & 12 & 30 & He$_{2}^{+}$ coinc. & {\scriptsize ICD}
                 & ~~$0$ % q
                 & $60.432 \pm 0.007$ % E_r
                 & $0.191 \pm 0.013$ % Gamma
                 & $0.158 \pm 0.006$ % s_d
                 & $-0.166 \pm 0.006$ % s_r
                 & $0.158 \pm 0.006$ % s_d + q^2 s_r
                 \\
    % electron_eKE_loss/fano_fits_he2_coinc/fit_Astrid_14_K_data.txt (total)
    {\scriptsize VMI} & 14 & 30 & He$_{2}^{+}$ coinc. & total
                 & $-3.01 \pm 0.26$ % q
                 & $60.359 \pm 0.011$ % E_r
                 & $0.214 \pm 0.012$ % Gamma
                 & $6.410 \pm 0.081$ % s_d
                 & $0.586 \pm 0.095$ % s_r
                 & $11.7 \pm 1.3$ % s_d + q^2 s_r
                 \\
    % electron_eKE_loss/fano_fits_he2_coinc/fit_Astrid_14_K_data.txt (EEL)
    {\scriptsize VMI} & 14 & 30 & He$_{2}^{+}$ coinc. & {\scriptsize EEL}
                 & $-4.54 \pm 0.84$ % q
                 & $60.379 \pm 0.018$ % E_r
                 & $0.236 \pm 0.020$ % Gamma
                 & $0.644 \pm 0.015$ % s_d
                 & $0.030 \pm 0.011$ % s_r
                 & $1.3 \pm 0.3$ % s_d + q^2 s_r
                 \\
    % electron_eKE_loss/fano_fits_he2_coinc/fit_Astrid_14_K_data.txt (ICD)
    {\scriptsize VMI} & 14 & 30 & He$_{2}^{+}$ coinc. & {\scriptsize ICD}
                 & ~~$0$ % q
                 & $60.455 \pm 0.011$ % E_r
                 & $0.187 \pm 0.022$ % Gamma
                 & $0.237 \pm 0.016$ % s_d
                 & $-0.256 \pm 0.016$ % s_r
                 & $0.237 \pm 0.016$ % s_d + q^2 s_r
                 \\
    % electron_eKE_loss/fano_fits_he2_coinc/fit_Astrid_16_K_data.txt (total)
    {\scriptsize VMI} & 16 & 30 & He$_{2}^{+}$ coinc. & total
                 & $-2.06 \pm 0.08$ % q
                 & $60.456 \pm 0.010$ % E_r
                 & $0.242 \pm 0.010$ % Gamma
                 & $2.175 \pm 0.027$ % s_d
                 & $0.528 \pm 0.037$ % s_r
                 & $4.4 \pm 0.2$ % s_d + q^2 s_r
                 \\
    % electron_eKE_loss/fano_fits_he2_coinc/fit_Astrid_16_K_data.txt (EEL)
    {\scriptsize VMI} & 16 & 30 & He$_{2}^{+}$ coinc. & {\scriptsize EEL}
                 & $-2.86 \pm 0.23$ % q
                 & $60.478 \pm 0.015$ % E_r
                 & $0.214 \pm 0.016$ % Gamma
                 & $0.202 \pm 0.005$ % s_d
                 & $0.034 \pm 0.005$ % s_r
                 & $0.48 \pm 0.06$ % s_d + q^2 s_r
                 \\
    % electron_eKE_loss/fano_fits_he2_coinc/fit_Astrid_16_K_data.txt (ICD)
    {\scriptsize VMI} & 16 & 30 & He$_{2}^{+}$ coinc. & {\scriptsize ICD}
                 & ~~$0$ % q
                 & $60.480 \pm 0.014$ % E_r
                 & $0.152 \pm 0.025$ % Gamma
                 & $0.127 \pm 0.013$ % s_d
                 & $-0.135 \pm 0.012$ % s_r
                 & $0.127 \pm 0.013$ % s_d + q^2 s_r
                 \\

%% file: table_1_d.tex
    % electron_eKE_loss/fano_fits_he3_coinc/fit_Astrid_10_K_data.txt (total)
    {\scriptsize VMI} & 10 & 30 & He$_{3}^{+}$ coinc. & total
                 & $-4.23 \pm 0.59$ % q
                 & $60.361 \pm 0.011$ % E_r
                 & $0.261 \pm 0.008$ % Gamma
                 & $0.042 \pm 0.001$ % s_d
                 & $0.002 \pm 0.001$ % s_r
                 & $0.08 \pm 0.02$ % s_d + q^2 s_r
                 \\
    % electron_eKE_loss/fano_fits_he3_coinc/fit_Astrid_10_K_data.txt (EEL)
    {\scriptsize VMI} & 10 & 30 & He$_{3}^{+}$ coinc. & {\scriptsize EEL}
                 & $-10.82 \pm 20.34$ % q
                 & $60.340 \pm 0.073$ % E_r
                 & $0.341 \pm 0.060$ % Gamma
                 & $0.003 \pm 0.001$ % s_d
                 & $0.000 \pm 0.000$ % s_r
                 & $0.003 \pm 0.001$ % s_d + q^2 s_r
                 \\
    % electron_eKE_loss/fano_fits_he3_coinc/fit_Astrid_10_K_data.txt (ICD)
    {\scriptsize VMI} & 10 & 30 & He$_{3}^{+}$ coinc. & {\scriptsize ICD}
                 & ~~$0$ % q
                 & $60.416 \pm 0.003$ % E_r
                 & $0.225 \pm 0.011$ % Gamma
                 & $0.000 \pm 0.000$ % s_d
                 & $-0.000 \pm 0.000$ % s_r
                 & $0.00023 \pm 0.00001$ % s_d + q^2 s_r
                 \\
    % electron_eKE_loss/fano_fits_he3_coinc/fit_Astrid_12_K_data.txt (total)
    {\scriptsize VMI} & 12 & 30 & He$_{3}^{+}$ coinc. & total
                 & $-2.34 \pm 0.11$ % q
                 & $60.412 \pm 0.007$ % E_r
                 & $0.228 \pm 0.008$ % Gamma
                 & $0.569 \pm 0.007$ % s_d
                 & $0.105 \pm 0.009$ % s_r
                 & $1.14 \pm 0.07$ % s_d + q^2 s_r
                 \\
    % electron_eKE_loss/fano_fits_he3_coinc/fit_Astrid_12_K_data.txt (EEL)
    {\scriptsize VMI} & 12 & 30 & He$_{3}^{+}$ coinc. & {\scriptsize EEL}
                 & $-2.29 \pm 0.29$ % q
                 & $60.468 \pm 0.021$ % E_r
                 & $0.239 \pm 0.022$ % Gamma
                 & $0.053 \pm 0.002$ % s_d
                 & $0.010 \pm 0.002$ % s_r
                 & $0.105 \pm 0.017$ % s_d + q^2 s_r
                 \\
    % electron_eKE_loss/fano_fits_he3_coinc/fit_Astrid_12_K_data.txt (ICD)
    {\scriptsize VMI} & 12 & 30 & He$_{3}^{+}$ coinc. & {\scriptsize ICD}
                 & ~~$0$ % q
                 & $60.440 \pm 0.007$ % E_r
                 & $0.191 \pm 0.013$ % Gamma
                 & $0.017 \pm 0.001$ % s_d
                 & $-0.018 \pm 0.001$ % s_r
                 & $0.017 \pm 0.001$ % s_d + q^2 s_r
                 \\
    % electron_eKE_loss/fano_fits_he3_coinc/fit_Astrid_14_K_data.txt (total)
    {\scriptsize VMI} & 14 & 30 & He$_{3}^{+}$ coinc. & total
                 & $-2.93 \pm 0.28$ % q
                 & $60.397 \pm 0.013$ % E_r
                 & $0.219 \pm 0.014$ % Gamma
                 & $0.775 \pm 0.011$ % s_d
                 & $0.073 \pm 0.013$ % s_r
                 & $1.40 \pm 0.16$ % s_d + q^2 s_r
                 \\
    % electron_eKE_loss/fano_fits_he3_coinc/fit_Astrid_14_K_data.txt (EEL)
    {\scriptsize VMI} & 14 & 30 & He$_{3}^{+}$ coinc. & {\scriptsize EEL}
                 & $-6.36 \pm 3.73$ % q
                 & $60.412 \pm 0.046$ % E_r
                 & $0.266 \pm 0.050$ % Gamma
                 & $0.066 \pm 0.004$ % s_d
                 & $0.002 \pm 0.002$ % s_r
                 & $0.15 \pm 0.12$ % s_d + q^2 s_r
                 \\
    % electron_eKE_loss/fano_fits_he3_coinc/fit_Astrid_14_K_data.txt (ICD)
    {\scriptsize VMI} & 14 & 30 & He$_{3}^{+}$ coinc. & {\scriptsize ICD}
                 & ~~$0$ % q
                 & $60.466 \pm 0.012$ % E_r
                 & $0.191 \pm 0.024$ % Gamma
                 & $0.026 \pm 0.002$ % s_d
                 & $-0.028 \pm 0.002$ % s_r
                 & $0.026 \pm 0.002$ % s_d + q^2 s_r
                 \\
    % electron_eKE_loss/fano_fits_he3_coinc/fit_Astrid_16_K_data.txt (total)
    {\scriptsize VMI} & 16 & 30 & He$_{3}^{+}$ coinc. & total
                 & $-1.95 \pm 0.09$ % q
                 & $60.488 \pm 0.012$ % E_r
                 & $0.249 \pm 0.012$ % Gamma
                 & $0.261 \pm 0.004$ % s_d
                 & $0.073 \pm 0.006$ % s_r
                 & $0.54 \pm 0.03$ % s_d + q^2 s_r
                 \\
    % electron_eKE_loss/fano_fits_he3_coinc/fit_Astrid_16_K_data.txt (EEL)
    {\scriptsize VMI} & 16 & 30 & He$_{3}^{+}$ coinc. & {\scriptsize EEL}
                 & $-1.61 \pm 0.17$ % q
                 & $60.588 \pm 0.025$ % E_r
                 & $0.205 \pm 0.026$ % Gamma
                 & $0.018 \pm 0.001$ % s_d
                 & $0.008 \pm 0.001$ % s_r
                 & $0.039 \pm 0.005$ % s_d + q^2 s_r
                 \\
    % electron_eKE_loss/fano_fits_he3_coinc/fit_Astrid_16_K_data.txt (ICD)
    {\scriptsize VMI} & 16 & 30 & He$_{3}^{+}$ coinc. & {\scriptsize ICD}
                 & ~~$0$ % q
                 & $60.492 \pm 0.014$ % E_r
                 & $0.151 \pm 0.025$ % Gamma
                 & $0.013 \pm 0.001$ % s_d
                 & $-0.013 \pm 0.001$ % s_r
                 & $0.013 \pm 0.001$ % s_d + q^2 s_r
                 \\

%% file: table_1_e.tex
    % electron_eKE_loss/elettra/fit_Elettra_14_K_data.txt (EEL)
    {\scriptsize HEA} & 14 & 50 & all $e^{-}$ & {\scriptsize EEL}
                 & $-2.94 \pm 0.15$ % q
                 & $60.405 \pm 0.010$ % E_r
                 & $0.223 \pm 0.011$ % Gamma
                 & $0.356 \pm 0.006$ % s_d
                 & $0.063 \pm 0.006$ % s_r
                 & $0.90 \pm 0.08$ % s_d + q^2 s_r
                 \\
    % electron_eKE_loss/elettra/fit_Elettra_14_K_data.txt (ICD)
    {\scriptsize HEA} & 14 & 50 & all $e^{-}$ & {\scriptsize ICD}
                 & $0.01 \pm 0.02$ % q
                 & $60.381 \pm 0.015$ % E_r
                 & $0.24 \pm 0.03$ % Gamma
                 & $0.133 \pm 0.008$ % s_d
                 & $-0.141 \pm 0.007$ % s_r
                 & $0.133 \pm 0.008$ % s_d + q^2 s_r
                 \\